\documentclass[letterpaper,twocolumn,showpacs,prl,aps,superscriptaddress]{revtex4}
\usepackage[latin1]{inputenc}
\usepackage{bm}
\usepackage{amsmath}
\usepackage{graphicx}
\usepackage{amssymb}

\begin{document}

\title{Non-Markovian quantum trajectories: an exact result}

\author{Angelo Bassi}
\email{bassi@ts.infn.it}
\affiliation{Dipartimento di Fisica Teorica,
Universit\`a di Trieste, Strada Costiera 11, 34014 Trieste, Italy.
 \\ Istituto Nazionale di Fisica Nucleare,
Sezione di Trieste, Strada Costiera 11, 34014 Trieste, Italy.}

\author{Luca Ferialdi}
\email{ferialdi@ts.infn.it}
\affiliation{Dipartimento di Fisica Teorica,
Universit\`a di Trieste, Strada Costiera 11, 34014 Trieste, Italy.
 \\ Istituto Nazionale di Fisica Nucleare,
Sezione di Trieste, Strada Costiera 11, 34014 Trieste, Italy.}

\begin{abstract}
We analyze the non-Markovian stochastic Schr\"odinger equation
describing a particle subject to spontaneous collapses in space (in
the language of collapse models), or subject to a continuous
measurement of its position (in the language of continuous quantum
measurement). For the first time, we give the explicit general
solution for the free particle case ($H = p^2/2m$), and discuss the
main properties. We analyze the case of an exponential correlation
function for the noise, giving a quantitative description of the
dynamics and of its dependence on the correlation time.
\end{abstract}


\maketitle

The theory of non-Markovian quantum dynamics is a subject of growing
interest, both from the theoretical point of view, as well as for
its experimental implications~\cite{soren}. On the more theoretical
side, interest ranges from the theory of open quantum
systems~\cite{1-4,bp,mail}, to the theory of continuous quantum
measurement~\cite{5-7}, to quantum
trajectories~\cite{891213,Diosi:97,Diosi:98}, to models of
spontaneous wave function
collapse~\cite{Pearle:96,Bassi:02,Adler:07,Adler:08}. With
particular reference to the latter, recent
investigations~\cite{Adler2:07} have shown that they might need to
be generalized to non-white noises, in order to be compatible with
current experimental constraints.

Unlike the theory of Markov dynamics in Hilbert space, which has
been deeply investigated and well understood, the theory of
non-Markovian quantum dynamics is still under construction.
Important results have been already obtained~\cite{bp,mail}. With particular
reference to stochastic Schr\"odinger equations (SSEs) in Hilbert
spaces, these have been formally generalized to non-Markovian
noises~\cite{Diosi:97}, but explicit results have been obtained only
for simple systems~\cite{Diosi:98}, or through approximation
schemes~\cite{Yu:99,Adler:08}.

In the Markovian case, among all SSEs, the following equation,
\begin{equation} \label{eq:white}
d \phi_t = \left[ - \frac{i}{\hbar} H dt + \sqrt{\lambda} q dW_t -
\frac{\lambda}{2} q^2 dt \right] \phi_t,
\end{equation}
has received considerable
attention~\cite{2021,222325,Chruscinski:92,Halliwell:95,Holevo:96,Bassi2:05,Bassi:08,Bassi5:08}.
$q$ is the position operator of the particle, $H$ its quantum
Hamiltonian, $W_t$ a standard Wiener process defined on a
probability space $(\Omega, {\mathcal F}, {\mathbb Q})$ and
$\lambda$ is a positive coupling constant~\cite{com1}. The evolution
described by Eq.~\eqref{eq:white} is manifestly non-unitary, the
non-Schr\"odinger terms being devised in order to reproduce the
collapse of the wave function~\cite{Ghirardi:86}.

The reason why Eq.~\eqref{eq:white} is so popular is that it
represents an excellent compromise between mathematical simplicity
and physical adequacy. From the mathematical point of view, it is
simple enough to be analyzed in great
detail~\cite{Holevo:96,Bassi2:05,Bassi:08}.{}~From the physical
point of view instead, it represents a very realistic model
describing a quantum particle subject to spontaneous collapses in
space (within collapse models~\cite{2021,Bassi2:05}), or a particle
whose position is continuously measured by an appropriate device
(within the theory of continuous quantum
measurement~\cite{222325,Chruscinski:92}), or a particle coupled to
an environment via its position (within the theory of open quantum
systems~\cite{Halliwell:95}). In all fields of applicability,
Eq.~\eqref{eq:white} has been used to get a deep insight into the
dynamics of more complicated physical situations.

It is then of primary interest to study the generalization of
Eq.~\eqref{eq:white} to the non-Markovian case. Such a
generalization has been first proposed in~\cite{Diosi:97}, and reads:
\begin{equation}\label{eq:freepart}
\frac{d}{dt} \phi_t = \left[ - \frac{i}{\hbar} H + \sqrt{\lambda} q
w_t - 2\sqrt{\lambda} q \int_0^t \!\!ds \, \alpha(t,s)
\frac{\delta}{\delta w_s} \right] \phi_t,
\end{equation}
where now $w_t$ is a Gaussian non-white noise defined on $(\Omega,
{\mathcal F}, {\mathbb Q})$, having zero average and the correlation
function $\alpha(t,s)$. The non-Markovian character is clearly
displayed by the third term, which depends on the whole past
history. For this reason, technically speaking, the integration
should begin at $s = - \infty$. Here we are making the assumption,
which actually is an approximation, that the state of the system at
time $0$ suffices to unfold the subsequent evolution. This would be
the case, e.g., if the system has reached an equilibrium
configuration which is independent of the way it has been reached,
and one is interested in studying what happens if at time 0 the
system is driven away from it by a sudden interaction.

Setting Eq.~\eqref{eq:freepart} has represented a very important
achievement. However it remains still somewhat formal, as no
explicit solutions are known. In this Letter we present a recent
result, whose technical details are reported in~\cite{Bassi:09}:
for the first time, the explicit expression
of the Green's function associated to Eq.~\eqref{eq:freepart} has
been computed, in the case of a free particle ($H = p^2/2m$), and
its properties have been analyzed in detail. The technique which has
been used can be straightforwardly generalized to include linear and
quadratic potentials (thus bounded systems can also be studied).
More complicated situations can be analyzed through a perturbation
expansion on $\sqrt{\lambda}$.

{\it The Green's function.} In~\cite{Diosi:97} it was first shown that the
Green's function $G(x,t;x_0,0)$ associated to
Eq.~\eqref{eq:freepart} allows for the following path-integral
representation:
\begin{equation}\label{eq:propform}
G(x,t;x_0,0) \; = \; \int^{q(t)=x}_{q(0)=x_0} \mathcal{D}[q] \;
e^{\mathcal{S}[q]} \,,
\end{equation}
where the `action' $\mathcal{S}[q]$, which is not standard, having
both a real and an imaginary part, is:
\begin{equation} \label{eq:action}
\mathcal{S}[q] = \int^t_0 \! ds \left[ \frac{i m}{2\hbar}\, q'^2_s +
\sqrt{\lambda} q_s w_s - \lambda q_s \!\int_0^t dr\,
\alpha(s,r)q_r\right].
\end{equation}
We have computed the path-integral in~\eqref{eq:propform} using the
polygonal approach of Feynman~\cite{Feynman:65,pirla}. The
calculation is long, in particular due to the last term which
contains a double integration reflecting the non-Markovian character
of the evolution; nevertheless the computation can be carried out
exactly. We report on the final result, focusing on the case of a
time-translation invariant noise ($\alpha(t,s) = \alpha(|t-s|)$),
which is sufficient for most physical purposes. In this case, the
Green's function becomes~\cite{Bassi:09}:
\begin{eqnarray}\label{eq:propts}
\lefteqn{G(x,t;x_0,0)=\sqrt{\frac{m}{2i\pi\hbar\, t\, u(t)}}} \nonumber\\
&&\cdot\exp\left[-\mathcal{A}_t (x_0^2+x^2)+\mathcal{B}_tx_0x
+\mathcal{C}_tx_0+\mathcal{D}_tx+\mathcal{E}_t\right].\qquad
\end{eqnarray}
The first two coefficients $\mathcal{A}_t$ and $\mathcal{B}_t$ are
deterministic functions of time and are defined as follows:
\begin{equation} \label{eq:matha}
\mathcal{A}_t = \frac{im}{2\hbar}\, f'_t(0), \qquad \mathcal{B}_t =
\frac{im}{\hbar}\, f'_t(t),
\end{equation}
while the remaining coefficients $\mathcal{C}_t$, $\mathcal{D}_t$
and $\mathcal{E}_t$ depend also on the noise $w_t$ through the
expressions:
\begin{eqnarray}
\mathcal{C}_t & = & -\frac{im}{2\hbar}\, h'_t(0) +
\frac{\sqrt{\lambda}}{2}\int_0^t dl\, w_l f_t(l),\\
\mathcal{D}_t & = & \frac{im}{2\hbar}\, h'_t(t) +
\frac{\sqrt{\lambda}}{2}\int_0^t dl\, w_l f_t(t-l),\\
\label{eq:mathe}\mathcal{E}_t&=&\frac{\sqrt{\lambda}}{2}\int_0^t
dl\, w_l h_t(l).
\end{eqnarray}
(Here above and in the following, the symbol $'$ denotes
differentiation with respect to the variable within parenthesis.)
The random function $h_t(s)$ satisfies the following non-homogeneous
integro-differential equation:
\begin{equation}\label{eq:hans}
\frac{im}{2\hbar} h''_t(s)+\lambda\int_0^t dr \alpha(s,r)h_t(r) =
\frac{\sqrt{\lambda}}{2}w_s\,,
\end{equation}
with boundary conditions $h_t(0)=h_t(t)=0$. The function $f_t(s)$
instead satisfies the homogeneous equation associated to
Eq.~\eqref{eq:hans}, with boundary conditions $f_t(0)=1$,
$f_t(t)=0$. Also the function $u(t)$ can be given an analytic
expression in terms of the solution of an integro-differential
equation; since however the whole square root in~\eqref{eq:propts}
represents a global factor whose real part looses importance when
normalizing the wave function, and whose imaginary part represents
an uninteresting global phase factor, we omit to write the explicit
expression of $u(t)$.

Eqs.~\eqref{eq:propts}-\eqref{eq:hans} represent our main result,
from which the subsequent discussion follows. One should notice the
quite remarkable fact that we have been able to compute the Green's
function associated to Eq.~\eqref{eq:freepart} (which can be applied
to any ${\mathcal L}^2$ initial state, giving its time evolution),
while in general non-Markovian dynamics do not allow for such a
thing as the Green's function. This fact is less surprising if one
looks back at how Eq.~\eqref{eq:freepart} was derived~\cite{891213}:
first the evolution was set by means of a propagator, and only
afterwards the associated differential equation was deduced.

Another relevant observation to make is that the structure of the
non-Markovian Green's function $G(x,t;x_0,0)$ is the same as the
corresponding Markovian one~\cite{Kolokoltsov:98,Bassi:08}, and of
course reduces to it in the white-noise limit, as proven
in~\cite{Bassi:09}. In particular, the exponent is quadratic in the
variables $x_0$, $x$, and the coefficients associated with the
quadratic terms do not depend on the noise. This fact has two
important consequences: first, the shape of Gaussian states is
preserved during the evolution; second, their spread evolves
deterministically in time. We will come back on these points later.
Since more general states can be written as superpositions of
Gaussian states, these facts suggest that any reasonable initial
state converges almost surely to a Gaussian state with a fixed
spread both in position as well as in momentum. This property holds
in the Markovian case, and has been subject of an intense
investigation~\cite{3435,Chruscinski:92,Halliwell:95,Bassi2:05,Bassi:08,Kolokoltsov:98}.
It would be important to check it also in a non-Markovian setting.

As a second relevant consequence of Eq.~\eqref{eq:propts}, one can
verify that the following ansatz:
\begin{equation} \label{eq:fdsfs}
\frac{\delta}{\delta w_s} \phi_t \; = \;
[q\,a_t(s)+p\,b_t(s)+c_t(s)]\phi_t,
\end{equation}
first proposed in~\cite{Diosi:98}, is correct; the three
coefficients have the following time dependence~\cite{Bassi:09}:
\begin{equation}
a_t(s) =  f_t(t-s) + \frac{f'_t(0)}{f'_t(t)}f_t(s),\quad b_t(s) \; =
\; \frac{1}{m}\frac{f_t(s)}{f'_t(t)},
\end{equation}
\vspace{-0.5cm}
\begin{equation}
c_t(s) = h_t(s) - \frac{f_t(s)}{2 f'_t(t)}\!\left(\! h'_t(t) +
\frac{i\sqrt{\lambda}\hbar}{m} \!\int_0^t
\!dl\,w_lf_t(t-l)\!\right)\!.
\end{equation}
One can then replace the functional derivative appearing
in~\eqref{eq:freepart} with~\eqref{eq:fdsfs}, giving the
non-Markovian equation a less cumbersome expression. The
form~\eqref{eq:fdsfs} for the functional derivative should make it
clear that the non-Markovian term of Eq.~\eqref{eq:freepart} depends
on the interplay between the Hamiltonian and the collapse terms,
since a term proportional to $p$ appears, which can come only from
the free part of the evolution. This is the ultimate reason why the
functional derivative can be computed explicitly only when all
operators appearing in Eq.~\eqref{eq:freepart} commute with each
other~\cite{Bassi:02}, or in simple enough cases like ours.

One can further prove~\cite{Bassi:09} that the mean position
${\mathbb E}_{\mathbb Q} \left[\langle q\rangle_t\right]$ and mean
momentum ${\mathbb E}_{\mathbb Q} \left[\langle p\rangle_t\right]$
evolve according to the classical laws. Moreover, the fluctuations
of the position of the particle around the average, measured by
$\mathbb{V}_q := \sqrt{\mathbb{E}_{\mathbb{Q}}\left[\langle
q\rangle_t- \mathbb{E}_{\mathbb{Q}}[\langle q\rangle_t]\right]^2}$,
scale with the inverse square root of its mass; this means that, the
bigger the system, the less random the motion within a given time
interval.

{\it Exponential correlation function.} The explicit form of the
coefficients ${\mathcal A}_t$--${\mathcal E}_t$ defining the Green's
function depend on the solution $h_t(s)$ of Eq.~\eqref{eq:hans} and
on the solution $f_t(s)$ of the corresponding homogeneous equation.
In general, this equation cannot be solved explicitly, though a
perturbation expansion is always possible, which gives meaningful
results to first orders in $\lambda$. Nevertheless, the solution can
be found for particular types of correlation
functions~\cite{Polyanin:08}. Among these, the physically most
meaningful example is the exponential correlation function:
\begin{equation}\label{eq:expcorr}
\alpha(t,s)=(\gamma/2)e^{-\gamma |t-s|}\,,
\end{equation}
where $\gamma$ is the inverse of the correlation time.

With this choice for $\alpha(t,s)$, the homogeneous equation for
$f_t(s)$ can be solved as follows. By differentiating twice
Eq.~\eqref{eq:hans} with $w_s = 0$, one can transform the
integro-differential equation into the fourth-order differential
equation~\cite{Bassi:09}:
\begin{equation}\label{eq:f4}
f''''(s)-\gamma^2f''(s)+i\gamma^2 \omega^2 f(s)=0\,,
\end{equation}
where $\omega=2\sqrt{\hbar\lambda/m}$. The general solution is
$f_t(s) = \sum_{k=1}^{2} [ f_{t,k} \sinh \upsilon_k s + g_{t,k}
\cosh \upsilon_k s ]$, where $f_{t,k}$, $g_{t,k}$ are determined by
the boundary conditions, and $\upsilon_1$, $\upsilon_2$ are the two
non-symmetric roots of the bi-quadratic characteristic polynomial
associated to Eq.~\eqref{eq:f4}:
\begin{equation} \label{eq:gdsfsdasda}
\upsilon_{1,2}=\sqrt{\left(\gamma^2\pm\zeta\right)/2}\,,
\qquad\zeta=\sqrt{\gamma^4-4i\gamma^2\omega^2}\,.
\end{equation}
Two boundary conditions are already given: $f_t(0) = 1$ and $f_t(t)
= 1$. The other two conditions can be recovered~\cite{Polyanin:08}
from the procedure which led to Eq.~\eqref{eq:f4} and read:
$f'''_t(0) = \gamma f''_t(0)$ and $f'''_t(t) = -\gamma f''_t(t)$.
Inserting these conditions, one obtains:
\begin{equation}\label{eq:fsol}
f_t(s) = \frac{\sum_k\left[r_t^k \sinh \upsilon_k(t-s) + u_t^k \cosh
\upsilon_k(t-s) - u_s^k\right]}{ \sum_{k}\left[2c + r_t^k \, \sinh
\upsilon_k t + u_t^k \cosh \upsilon_k t\right]},
\end{equation}
with $k = 1,2$ and where $r_t^k = a_{\bar{k}} \cosh
\upsilon_{\bar{k}}t +b_{\bar{k}}\sinh \upsilon_{\bar{k}}t$ and
$u_t^k = d_k\sinh \upsilon_{\bar{k}}t-c \cosh \upsilon_{\bar{k}}t$;
we have also defined: $a_k = \gamma
\upsilon_k^3[\upsilon_k^2+(-1)^{\bar{k}} \zeta]$, $b_k
=\upsilon_k^2[\upsilon_k^4+(-1)^{\bar{k}}\gamma^2\zeta]$, $c =
\upsilon_1^3\upsilon_2^3$, $d_k = -\gamma
\upsilon_k^3\upsilon_{\bar{k}}^2$, with $\bar{k}=2$ if $k=1$,
$\bar{k}=1$ if $k=2$.

The function $h_t(s)$ can be found in a similar way, though its
expression is more complicated, as $h_t(s)$ solves the whole
inhomogeneous equation. Taking into account the boundary conditions,
$h_t(s)$ takes the form: $h_t(s) = h^{\text{\tiny P}}_t(s) -
h^{\text{\tiny P}}_t(t) f_t(t-s)$, where $h^{\text{\tiny P}}_t(s)$
is a particular solution of~\eqref{eq:hans}, namely:
\begin{eqnarray}\label{eq:hp}
h^{\text{\tiny P}}_t(s) & = &
-\frac{i\sqrt{\lambda}\hbar}{m}\int_0^s
\bar{f}_s(l)\left(w''_l-\gamma^2w_l\right)dl\,, \nonumber\\
\bar{f}_s(l)& = & \frac{\sinh
\upsilon_1(s-l)}{\upsilon_1}-\frac{\sinh
\upsilon_2(s-l)}{\upsilon_2}\,.
\end{eqnarray}
The problem has been completely solved. One can check that in the
white-noise limit $\gamma\rightarrow\infty$ ($\alpha(t,s)
\rightarrow \delta(t-s)$), one recovers the well-known Markovian
expressions.

{\it Evolution of Gaussian states.} The analysis of Gaussian states
is particularly useful in order to understand the behavior of a
typical physical state. As previously anticipated, the shape of
Gaussian wave functions does not change in time. In fact, an initial
state:
\begin{equation}
\phi_0(x)=\exp[-\alpha_0 x^2+\beta_0x+\gamma_0]\,,
\end{equation}
preserves its functional dependence on $x$, while the complex
parameters $\alpha_0$, $\beta_0$ and $\gamma_0$ evolve in time as
follows:
\begin{eqnarray}
\alpha_t&=&\mathcal{A}_t-\frac{\mathcal{B}^2_t}{4(\alpha_0+\mathcal{A}_t)},
\quad
\beta_t = -\frac{\mathcal{C}_t+\beta_0}{4(\alpha_0+\mathcal{A}_t)}+\mathcal{D}_t \nonumber \\
\gamma_t&=&\gamma_0+\mathcal{E}_t+\frac{(\mathcal{C}_t+\beta_0)^2}{4(\alpha_0+\mathcal{A}_t)}\,.
\end{eqnarray}
Analyzing the above expressions with the help of
Eqs.~\eqref{eq:matha}--\eqref{eq:mathe}, one immediately sees that
the evolution of $\alpha_t$ is deterministic, while $\beta_t$ and
$\gamma_t$ have stochastic terms. This means that, like in the
white-noise case, both the spread in position and in momentum of
$\phi_t(x)$, which are given by $\alpha_t$, evolve deterministically
in time. On the other hand, both the mean position and the mean
momentum, which depend both on $\alpha_t$ and $\beta_t$, have
stochastic components; their stochastic averages instead evolve
according to classical laws, as we have already anticipated.

We focus now our attention on the spread in position $\sigma(t) =
1/2\sqrt{\alpha^{\text{\tiny R}}_t}$, in the case of the exponential
correlation function treated before.
\begin{figure}[t!]
\begin{center}
\includegraphics[width=8.5cm]{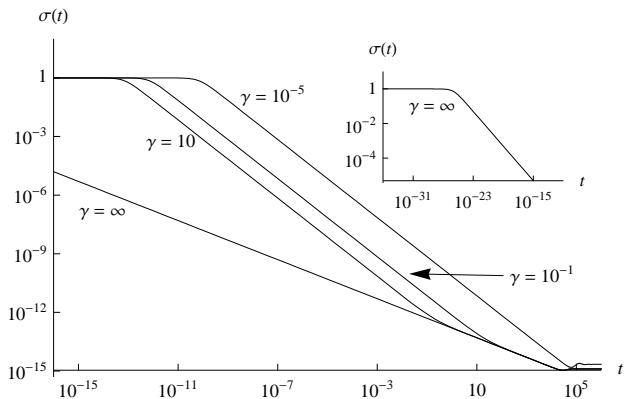}
\caption{Time evolution, under the assumption of an exponential
correlation function, of the spread in position $\sigma(t)$ of a
Gaussian wave function. $\sigma(0)$ has been set = 1 m. The value
$\gamma=\infty$ corresponds to the Markovian case. The other
parameters have been chosen as follows: $m = 1$ Kg, $\lambda_0 =
10^{-2}$ m$^{-2}$ sec$^{-1}$. Time is measured is sec, distances in
m.} \label{fig:1}
\end{center}
\end{figure}
Fig.~\ref{fig:1} shows how the spread evolves, for different values
of $\gamma$. Qualitatively the behavior is the same for any
$\gamma$: the wave function shrinks in space, reaching an asymptotic
finite value. On a more quantitative level, we see that the stronger
$\gamma$, the faster the collapse. One can also notice that the
collapse is effective starting with relatively small values of
$\gamma$: a value $\gamma \sim 10$ sec$^{-1}$ already ensures that
after about $10^{-3}$ sec the wave function has collapsed below
$10^{-5}$ cm, which is the threshold chosen by
GRW~\cite{Ghirardi:86}, below which a state can be considered as
localized. This means that the possibility opens for non-Markovian
models to be as effective as the corresponding white-noise models as
far as the collapse process is concerned, but, at the same time, to
give different physical predictions regarding specific experimental
situations. This possibility has first been suggested
in~\cite{Adler2:07}.

{} From the previous expressions one can explicitly compute the
asymptotic value of $\alpha_t$, which is:
\begin{equation}
\alpha_{\infty} = \lim_{t\rightarrow\infty}\alpha_t=-\frac{i
m}{2\hbar}(\upsilon_1+\upsilon_2-\gamma)\,.
\end{equation}
The quantity $1/2\sqrt{\alpha^{\text{\tiny R}}_{\infty}}$ is the
final spread in position to which all Gaussian states (and,
reasonably, any initial state) converge to, in the long-time limit.

{\it Conclusion.} We have computed for the first time the Green's
function associated to the motion of a free particle as described by
Eq.~\eqref{eq:freepart}, from which the entire non-Markovian
dynamics can be unfolded. We have analyzed the physically important
case of an exponential correlation function. By studying Gaussian
states, we have seen how the collapse occurs, and have derived an
exact expression for the asymptotic spread. The tools we have
employed to derive the above results are flexible and can be applied
to more complex physical situations.

{\it Acknowledgements.} We wish to thank S.L. Adler, A. Fonda and G.C. Ghirardi
for many useful conversations.

\def\polhk#1{\setbox0=\hbox{#1}{\ooalign{\hidewidth
  \lower1.5ex\hbox{`}\hidewidth\crcr\unhbox0}}} \def\cprime{$'$}
  \def\cprime{$'$}


\begin{thebibliography}{26}
\bibitem{com1} Often, a
non-linear SDE is considered together with, or in place of,
Eq.~\eqref{eq:white}; we will not enter here into the relationship
between the two equations.
\bibitem{pirla}
The same result can be obtained using the standatd operator
formalism for non-Markovian collapse dynamics first proposed
in~\cite{Pearle:96}, in place of the path-integral formalism. We
thank P. Pearle for valuable comments on this issue.
\end{thebibliography}
\end{document}